\documentclass[reprint,prl,amsmath,amssymb,aps]{revtex4-2}
\usepackage[utf8]{inputenc}
\usepackage{amsmath}
\usepackage{amsthm}
\usepackage{amssymb}
\usepackage{graphicx}

\makeatletter

\DeclareTextSymbolDefault{\textquotedbl}{T1}

\usepackage{amsthm}

\usepackage{dcolumn}
\usepackage{bm}
\usepackage{braket}
\usepackage{amsthm}

\newcommand{\dscf}{$\Delta\mathrm{SCF}$ }

\makeatother

\begin{document}
\title{Occupancy Extrapolation: Reaching Many Excited Electronic States from
Ground State Calculations}
\author{Yichen Fan}
\affiliation{Department of Chemistry, Duke University, Durham, North Carolina 27708}
\author{Weitao Yang}
\email{weitao.yang@duke.edu}

\affiliation{Department of Chemistry and Department of Physics, Duke University,
Durham, North Carolina 27708}
\begin{abstract}
The $\Delta$SCF DFT approach defines the system energy as a function of orbital occupancy. Inspired by Landau Fermi liquid theory, we develop an occupancy extrapolation (OE) method that captures excited-state energies via a Taylor expansion of the energy with respect to occupation fluctuation from a reference state. OE retains the physics of $\Delta$SCF while offering a physical interpretation of excitation energies as sums of quasiparticle energies and their generalized screened interactions.  It yields accurate valence, Rydberg, and charge-transfer excitation energies at $O(N^3)$ cost, avoids separate SCF calculations for each excited state, and enables efficient large-scale excited-state simulations from ground-state calculations.

\end{abstract}
\maketitle
Electronic excitations are fundamental to understanding photophysical
and photochemical processes and are key quantities in spectroscopy,
optoelectronics, and materials design. To model electronic excitations, a wide range of theoretical frameworks
has been developed over the past decades.\cite{geritbrandenburgNewApproachesStudy2020} The symmetry-adapted cluster configuration
interaction (SAC-CI)\cite{NakatsujiClusterTheory1978,nakatsujiClusterExpansionWavefunction1979} and equation-of-motion coupled-cluster
(EOM-CC) \cite{monkhorstCalculationPropertiesCoupledcluster1977, bartlettCoupledclusterTheoryIts2012, bartlettPerspectiveCoupledclusterTheory2024} are widely used approaches for excited state calculations built upon exponential cluster expansions of the wave function. 
Alternatively, time-dependent density functional theory (TD-DFT)\cite{rungeDensityFunctionalTheoryTimeDependent1984,casida1995timeDependentMolecules,bauernschmittTreatmentElectronicExcitations1996c} extends Kohn-Sham density functional theory (KS-DFT)
\cite{kohnSelfConsistentEquationsIncluding1965} to excited states through linear response. Beyond TD-DFT, the random phase approximation (RPA)\cite{thoulessVibrationalStatesNuclei1961, dunningNonempiricalCalculationsExcited1967} describes excitations through the particle-hole linear response, and the particle-particle random phase approximation (pp-RPA) \cite{yangDoubleRydbergCharge2013a, yangExcitationEnergiesParticleparticle2014a,liParticleParticleRandomPhase2024} formulates excitations as two-electron attachment to an $N-2$ reference system, which aims to capture physics with strong static correlation or double-excitation character. 
Beyond these approaches, the Bethe-Salpeter equation (BSE)
formalism \cite{salpeterRelativisticEquationBoundState1951,shamManyParticleDerivationEffectiveMass1966,strinatiApplicationGreensFunctions1988a,rohlfingElectronHoleExcitationsSemiconductors1998,onida_electronic_2002, blaseBetheSalpeterEquation2018, blaseBetheSalpeterEquation2020a, koskeloShortRangeExcitonicPhenomena2025}
provides a Green's Function based description of excitations, typically built upon quasiparticle energies obtained from the GW approximation\cite{hedinNewMethodCalculating1965},
while explicitly accounting for electron–hole interactions.

In contrast to these approaches, the $\Delta$ self-consistent
field ($\Delta$SCF) \cite{slaterStatisticalExchangeTotal1970c}
models excitations by adopting non-Aufbau electron configurations
and optimizes orbitals directly. The \dscf
method often provides a good estimation of excited state properties
including excitation energies \cite{haitExcitedStateOrbital2020}
and oscillator strength.\cite{bourneworsterReliableTransitionProperties2021} As DFT was originally formulated for ground state, the theoretical foundation \dscf had been unclear, despite the long history of applications\cite{vandaeleDSCFMethodNonadiabatic2022a}. Recent work establishes that one can use the defining variable of a noninteracting reference system, the density matrix for example, to formulate the universal energy functional with its minimum as the energy for the ground state and its stationary points as the energies of excited states, providing rigorous foundation for the \dscf approach. 

\dscf is frequently used with notable success in computational
spectroscopy, yet it has important drawbacks.\cite{gunnarssonExchangeCorrelationAtoms1976,jonesDensityFunctionalFormalism1989,gilbertSelfConsistentFieldCalculations2008a, malisOriginSingletExcited2026}
(1) The \dscf calculations are frequently troubled by SCF convergence. 
As the \dscf is evaluating the wave functions for excited states, they may collapse to the ground states during the variational orbital optimization. (2) \dscf relies on a state-specific non-Aufbau electron configuration, and multiple SCF cycles are necessary, when multiple excited states are needed, which introduces extra computational cost. (3) As the \dscf constructs 
an effective one-particle Hamiltonian, it is inherently
a single-determinant method and may break both spin symmetry for singlet excitations and some spatial symmetries, such as the $\pi$ symmetry in diatomic molecules.
For applications to complex systems and in a large scale, it would be desirable  
to leverage the accuracy of the $\Delta\mathrm{SCF}$ approach without
the additional challenging individual SCF calculations for each excited
state. 

A first and inspiring approach is the semiphenomenological Landau Fermi Liquid
(LFL) Theory, developed in the late 1950s for weakly interacting
homogeneous systems \cite{l.d.TheoryFermiLiquid1957,l.d.TheoryFermiLiquid1959,LandauFermiLiquidTheory1991,negele_quantum_2018}.
In LFL, low-lying excitations are modeled as small perturbations in
the fluctuation of the occupation numbers in the plane waves near the Fermi surface.
The corresponding excitation energy is expressed as a Taylor expansion:
\begin{equation}
\delta E=\sum_{\vec{k}}\varepsilon_{\vec{k}}^0\delta n(\vec{k})+\tfrac{1}{2\mathcal{V}}\sum_{\vec{k},\vec{k'}}f(\vec{k},\vec{k'})\delta n(\vec{k})\delta n(\vec{k'}),\label{LFL}
\end{equation}
where $\delta n(\vec{k})$ denotes a small change in the occupation
number, and the expansion coefficients
$\varepsilon_{\vec{k}}^{0}$ are the energies of the corresponding non-interacting system and $f(\vec{k}, \vec{k'})$ is the interaction energy of quasiparticles of momentum $\vec{k}$ and $\vec{k'}$. LFL theory has not been used for specific excited states, but has been successfully applied to the macroscopic behavior of a normal Fermi liquid at low temperatures. 

The LFL theory describes low-energy excitations within a uniform system; however, extending it to inhomogeneous systems introduces additional complexity; the energy is expanded in terms of the 1-particle density matrix instead of just orbital occupations. \cite{chakravartyDynamicResponseInhomogeneous1979,LandauFermiLiquidTheory1991,nozieres1999theory}

For finite molecular systems, several approaches have been developed to achieve Landau-Fermi-liquid-like
expansions of excited-state energies from ground-state quantities. 
Slater's transition-state
method \cite{slaterNonintegralOccupationNumbers1969,slaterStatisticalExchangeTotal1970c,slaterStatisticalExchangeCorrelationSelfConsistent1972a}
casts excitation energies as a power series in orbital occupations, with orbital relaxation
incorporated by self-consistent calculations at fractional occupations; subsequent refinements
\cite{williamsGeneralizationSlatersTransition1975,hirao_improved_2021,janaSlaterTransitionMethods2023}
improve accuracy by sampling multiple fractional points and higher-order fits. 

Related ideas appear in ensemble density
functional theory  \cite{theophilouEnergyDensityFunctional1979,grossRayleighRitzVariationalPrinciple1988,grossDensityfunctionalTheoryEnsembles-I1988,oliveiraDensityfunctionalTheoryEnsembles-II1988,filatovEnsembleDFTApproach2016}, which provides an analytic route to excitation energies from ensemble variational principles, but its
practical accuracy is limited by approximate ensemble functionals and spurious ``ghost''
contributions. Constricted variational DFT (CV-DFT) \cite{zieglerRelationTimedependentVariational2009a,zieglerCalculationChargeTransfer2010,zieglerImplementationSelfconsistentConstricted2012}
is an analytic expansion of the energy in terms of the density matrix, 
yet it typically requires solving additional constrained/self-consistent equations for targeted
excitations to capture relaxation and screening effects. These requirements motivate methods that
retain the occupation-expansion viewpoint while avoiding state- or ensemble-specific self-consistent
calculations.

Indeed, for the energy of a single quasiparticle describing a charged $(N\pm1)$ excitation of an N electron system, a direct and very simple method has been developed, called quasiparticle energies from the density functional theory (QE-DFT)\cite{meiApproximatingQuasiparticleExcitation2018}. QE-DFT calculates the $(N-1)/(N+1)$ excitation energy just from the corresponding occupied/virtual orbital energy of the N-electron system in its ground state. The theoretical foundation for the QE-DFT approach has been recently established in the general chemical potential theorem–the noninteracting one-electron orbital energies in DFT ground states and \dscf excited states correspond to chemical potentials of electron addition or removal, from an N -particle ground or excited state to an $(N\pm1)$-particle ground or excited state\cite{yangOrbitalEnergiesAre2024}. This result establishes the physical meaning of orbital energies as approximation to the corresponding excited-state ionization potentials and electron affinities, for both ground and excited states of a molecule or a bulk system. For the approximation to be accurate, the functional used should have minimal delocalization error\cite{mori-sanchezLocalizationDelocalizationErrors2008,  meiApproximatingQuasiparticleExcitation2018, liLocalizedOrbitalScaling2018, suPreservingSymmetryDegeneracy2020,yangOrbitalEnergiesAre2024}. Sometimes, the systematic delocalization error can also be partially canceled by using QE-DFT for neutral excitation energies from the orbital energy differences, leading to accurate results\cite{meiApproximatingQuasiparticleExcitation2018, meiChargeTransferExcitation2019, meiExcitedStatePotentialEnergy2019}. The theoretical foundation for using orbital energy to approximate excited-state quasiparticle energy seamlessly connects with that for the use of highest occupied molecular orbital (HOMO) and lowest unoccupied molecular orbital (LUMO) energies to approximate the ground state ionization energy and electron affinity\cite{janakProofThatfracensuremathpartialEensuremathpartialn_iensuremathepsilon1978,perdewDensityFunctionalTheoryFractional1982a,perdewPhysicalContentExact1983a,perdewCommentSignificanceHighest1997a, cohenFractionalChargePerspective2008, mori-sanchezDiscontinuousNatureExchangeCorrelation2009, yangDerivativeDiscontinuityBandgap2012}.

In the present work, we seek to generalize the QE-DFT approach to describe any 
quasiparticle excitations directly from the neutral ground state calculations.
We start with the excited DFT formulation\cite{yangFoundationDeltaSCFApproach2024} 
based on a universal energy functional $E_v[\gamma_s(\mathbf{x}, \mathbf{x'})]$ for
a physical system in an external potential $v$, using the noninteracting 1-particle
density matrix $\gamma_s(\mathbf{x}, \mathbf{x'})$ as the basic variable. 
$\gamma_s(\mathbf{x}, \mathbf{x'})$ can be expressed in terms of spin orbitals 
$\phi_{p\sigma}$ and occupation $n_{p\sigma}$. 
\begin{equation}
\gamma_{s}({\bf x},{\bf x}')=\sum_{q\sigma}n_{q\sigma}\phi_{q\sigma}(\mathbf{x})\phi_{q\sigma}^{*}({\bf x}').
\end{equation}
For the ground state, its energy is the minimum of $E_{v}\left[\gamma_{s}({\bf x},{\bf x}')\right]$
with respect to $\gamma_{s}({\bf x},{\bf x}')$, resulting in an
Aufbau occupation for $\left\{ n_{q\sigma}^0\right\} $. For excited
states, the energy is a stationary point of $E_{v}\left[\gamma_{s}({\bf x},{\bf x}')\right]$
with respect to $\gamma_{s}({\bf x},{\bf x}')$, leading to a non-Aufbau
$\left\{ n_{q\sigma}\right\} $\cite{yangFoundationDeltaSCFApproach2024}. Thus, we obtain
the system energy depending on the occupation $\left\{ n_{q\sigma}\right\} $
that specifies the corresponding state. Alternatively, as in
the practice of $\Delta$SCF, one can first specify a set of occupations
$\left\{ n_{q\sigma}\right\} $ and requires the energy be stationary with respect
to the orbitals to reach the desired ground or excited state. Therefore,
we define the following energy function:
\begin{equation}
E_{v}(\left\{ n_{q\sigma}\right\} )=\mathrm{Stat}_{\phi_{q\sigma}}E_{v}\left[\gamma_{s}({\bf x},{\bf x}')\right],
\end{equation}
which is an optimization under the constraint of a given $\left\{ n_{q\sigma}\right\} $.
This constrained optimization uniquely leads to the desired
 state, ground or excited, described by $\left\{ n_{q\sigma}\right\}$ that in turn defines the excitation quantum number $n_s$ of the noninteracting reference system in the formulation of $\Delta$SCF \cite{yangFoundationDeltaSCFApproach2024}.
\begin{figure}[]
\centering \includegraphics[width=.8\linewidth]{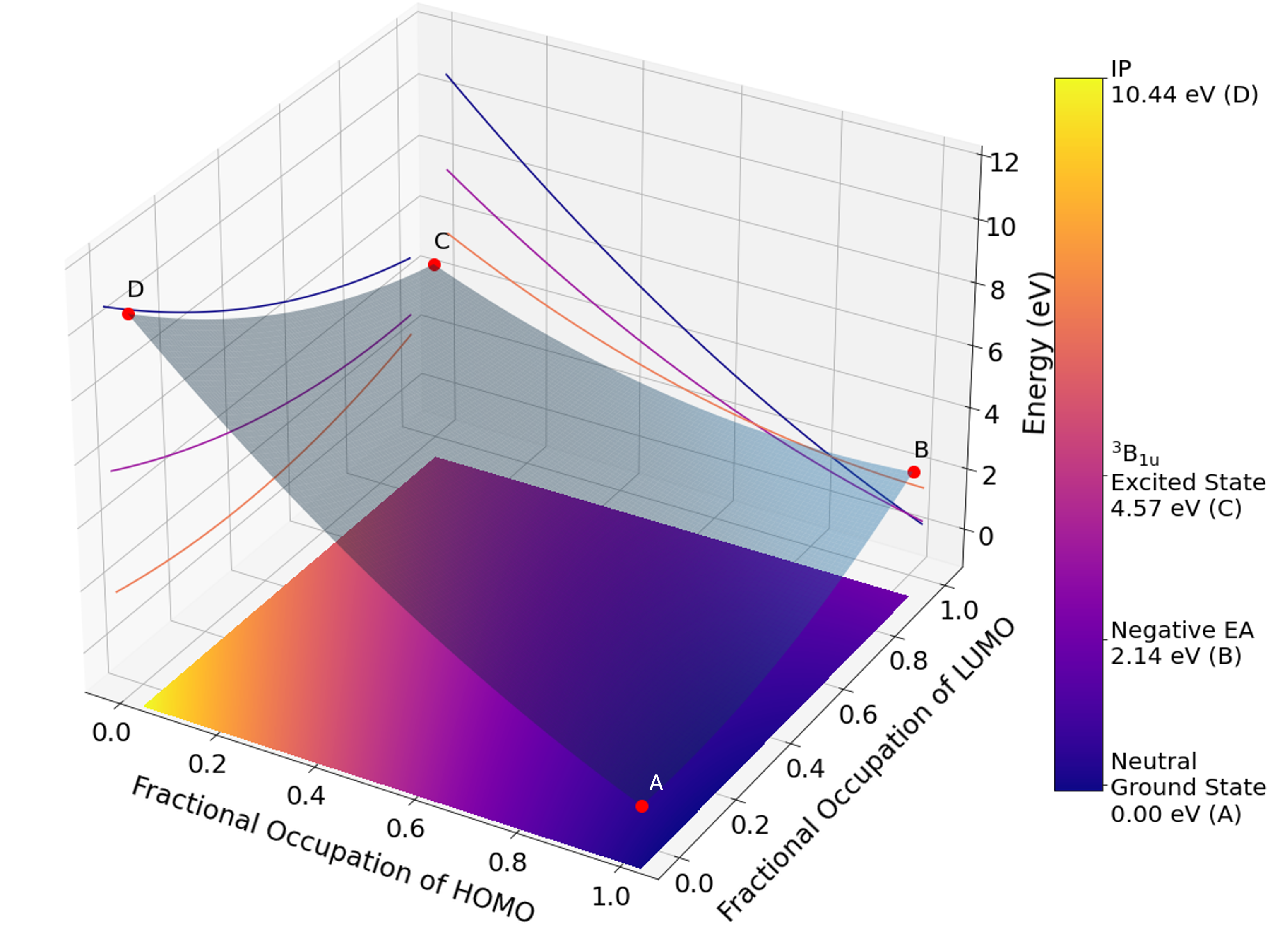} 
\caption{Occupancy extrapolation (OE@BLYP) energy surface for ethylene ($\rm{C_2H_4}$) as a function of fractional occupations of the HOMO and LUMO, computed with the BLYP functional\cite{beckeDensityfunctionalExchangeenergyApproximation1988,leeDevelopmentColleSalvettiCorrelationenergy1988}. Points A–D denote key electronic states: A, neutral ground state (0 eV); B, N+1 ground state (-EA = 2.14 eV); C, neutral triplet excited state $^{3}\mathrm{B}_{1u}$ (4.57 eV); D, N-1 ground state (IP = 10.44 eV). The color map represents total energy (yellow: higher, blue: lower). Convex energy behavior along A-B, A-D, C-B, and C-D reflects ground- and excited-state delocalization error of BLYP; they should be all straight lines for the exact functional \cite{perdewDensityFunctionalTheoryFractional1982a, yangFractionalChargesLinear2024b} The A-C convexity is physical and reflects particle-hole interaction.}

\label{fig:oe-surface} 
\end{figure}

The energy function $E_{v}(\left\{ n_{q\sigma}\right\} )$ defines
the physical ground state and excited state energies at discrete points
of integer occupation numbers (corresponding to a determinant). A
continuous extension in Eq. (3) can be made with the same definition using fractional occupations for zero temperature \cite{janakProofThatfracensuremathpartialEensuremathpartialn_iensuremathepsilon1978}
which can be viewed as using a $\gamma_{s}$ that is from an ensemble
of determinants, or as in finite temperature LFL theory \cite{LandauFermiLiquidTheory1991}. Figure \ref{fig:oe-surface} shows $E_{v}(\left\{ n_{q\sigma}\right\} )$
as a function of the HOMO and LUMO occupations. Special ensembles
with just a single fractional occupation are the fractional charge
states for ground states\cite{perdewDensityFunctionalTheoryFractional1982a} and for excited states\cite{yangFractionalChargesLinear2024b}. They have special physical meaning:
they arise in the dissociation limit of molecular systems \cite{yangDegenerateGroundStates2000,yangFoundationDeltaSCFApproach2024}. The exact
energy functional $E_{v}\left[\gamma_{s}({\bf x},{\bf x}')\right]$
is a linear line interpolating the energies at integers as a function
of the fractional charges, expressed as a linear combination of the
$\gamma_{s}$ of the $N$\textminus electron and $\gamma_{s}$  of the $(N+1)$\textminus electron
excited state. This is related to the concepts of excited-state chemical
potentials, which describe the slopes of these linear lines\cite{yangOrbitalEnergiesAre2024}. 

Having the energy now as a continuous function of occupations, we
can use a Taylor expansion in terms of occupation fluctuations $\delta n_{q\sigma}=n_{q\sigma}-n_{q\sigma}^{0}$
to extrapolate to the energy of an excited state described by occupation
$\left\{ n_{q\sigma}\right\} $, starting from a given reference state
$\left\{ n_{q\sigma}^{0}\right\} $, usually the ground state. Our
OE expansion is the following:
\begin{align}
&E_{v}(\left\{ n_{q\sigma}\right\} ) -E_{v}(\left\{ n_{q\sigma}^{0}\right\} )
=\sum_{q}\left.\tfrac{\partial E_{v}}{\partial n_{q\sigma}}\right|_{n_{q\sigma}^{0}}\delta n_{q\sigma}\nonumber\\
&+\tfrac{1}{2}\sum_{qr}\left.\tfrac{\partial^{2}E_{v}}{\partial n_{q\sigma}\partial n_{r\sigma}}\right|_{n_{q\sigma}^{0}}\delta n_{q\sigma}\delta n_{r\sigma}+O(\delta n_{q\sigma}^{3})
\label{eq:OE_Energy}
\end{align}
where the sum includes all quasiparticles that have a nonzero occupation
deviation $n_{q\sigma}-n_{q\sigma}^{0}$.

The first-order derivatives $\partial E_v/\partial n_{q\sigma}$ in
Eq.~(\ref{eq:OE_Energy}) correspond to the ground-state KS eigenvalues
for continuous functionals of the electron density
\cite{janakProofThatfracensuremathpartialEensuremathpartialn_iensuremathepsilon1978}
or to the GKS eigenvalues for continuous functionals of the density
matrix $\gamma_s$
\cite{cohenFractionalChargePerspective2008}:
\begin{equation}
\left.\tfrac{\partial E_{v}}{\partial n_{p\sigma}}\right|_{\{ n_{p\sigma}^{0}\}}
=\varepsilon_{p\sigma},
\label{eq:poor_qp_estimation}
\end{equation}
where $\varepsilon_{p\sigma}$ is the (G)KS eigenvalue of orbital
$\phi_{p\sigma}$. The second-order derivatives describe the interaction
between quasiparticles. The diagonal second derivatives were first derived
in Ref.~\cite{yangAnalyticalEvaluationFukui2012}, and the complete
second derivatives can be obtained in many ways
\cite{meiExactSecondOrderCorrections2021a}:
\begin{align}\label{kappa}
&\left.\tfrac{\partial^2E_v}{\partial n_{p\sigma}\partial n_{q\tau}}\right|_{\{n^0_{p\sigma}\}}
=\kappa_{p\sigma q\tau} \nonumber\\
&=(\phi_{p\sigma}\phi_{p\sigma}^*|K^{\sigma\tau}+\sum_{\sigma\upsilon}K^{\sigma\sigma}\chi^{\sigma\upsilon}K^{\upsilon\tau}|\phi_{q\tau}\phi_{q\tau}^*)
\end{align}
where $K^{\sigma\tau}$ is the spin resolved Hartree–exchange–correlation kernel, defined as the
functional derivative of the spin (G)KS Hamiltonian with respect to the
spin density matrix $\gamma_s$, and $\chi^{\sigma\upsilon}$ is the spin density response
function.

Further details of Eq. (\ref{kappa}) are provided in the Supplemental Material (SM)\cite{SM}. As noted previously, $K+K\chi K$ is the 4-point generalized screened interaction \cite{meiExactSecondOrderCorrections2021a}. Under the approximation of ignoring exchange-correlation contribution in $K$ and taking the RPA approximation instead of the DFT linear response for $\chi$, $K+K\chi K$ would be equal to the static screened Coulumb interaction used in GW and BSE theory.

Beyond excitation energies, we can also obtain orbital energies of an excited state just from ground state calculations by taking the derivative of  the QE expansion. Eq. (\ref{eq:OE_Energy}) with respect to orbital occupation:

\begin{equation}
\text{\ensuremath{\varepsilon_{q\sigma}^{*}}=}\left.\frac{\partial E_{v}}{\partial n_{q\sigma}}\right|_{\left\{ n_{p\sigma}\right\} }=\varepsilon_{q\sigma}+\sum_{qr}\kappa_{qr}\delta n_{q\sigma}.
\label{eq:ExOrbEnergy}
\end{equation}

Beyond the computational advance, OE also offers a physical interpretation.
We use a quasiparticle for adding an electron to the
unoccupied orbital $\phi_{a\sigma}$, for which $\delta n_{a\sigma}$=$1$,
and a quasihole for taking an electron from an occupied orbital $\phi_{i\sigma}$, in which $\delta n_{i\sigma}$=$-1$. We reorganize the second order
OE expansion, Eq. (\ref{eq:OE_Energy}), for an excited state with
$N_{p}$ quasiparticles and $N_{h}$ quasiholes as
\begin{widetext}
\begin{align}
\Delta E_{v}^{N_{p}N_{h}} &=\sum_{i}^{N_{h}}\varepsilon_{i\sigma}\delta n_{i\sigma}+\sum_{a}^{N_{p}}\varepsilon_{a\sigma}\delta n_{a\sigma}+\frac{1}{2}\left(\sum_{ij}^{N_{h}}\kappa_{i\sigma j\tau}\delta n_{i\sigma}\delta n_{j\tau}
 +\sum_{ab}^{N_{p}}\kappa_{a\sigma b\tau}\delta n_{a\sigma}\delta n_{b\tau} 
 + 2\sum_{ia}^{N_{p}N_{h}}\kappa_{i\sigma b\tau}\delta n_{i\sigma}\delta n_{b\tau}\right)\nonumber \\
&= -\sum_{i}^{N_{h}}\varepsilon_{i\sigma}^{QP}+\sum_{a}^{N_{p}}\varepsilon_{a\sigma}^{QP}+\frac{1}{2}\left(\sum_{i\sigma\neq j\tau}^{N_{h}}\kappa_{i\sigma j\tau}+\sum_{a\sigma\neq b\tau}^{N_{p}}\kappa_{a\sigma b\tau}-2\sum_{i\sigma, a\tau}^{N_{p}N_{h}}\kappa_{i\sigma a\tau}\right),
 \label{eq:OE_interpretation}
\end{align}
\end{widetext}
where we use $\delta n_{a\sigma}=1$, $\delta n_{i\sigma}=-1$,
and define the ground state quasihole and quasiparticle energy $\varepsilon_{i\sigma}^{QP}$ and $\varepsilon_{a\sigma}^{QP}$
as 
\begin{equation}
\varepsilon_{i\sigma}^{QP}=\varepsilon_{i\sigma}-\tfrac{1}{2}\kappa_{i\sigma i\sigma};\; \varepsilon_{a\sigma}^{QP}=\varepsilon_{a\sigma}+\tfrac{1}{2}\kappa_{a\sigma a\sigma}.
\label{eq:QPpm}
\end{equation}
Thus, the excitation energy, Eq. (\ref{eq:OE_interpretation}), has
a clear physical picture: it is the sum of the energy difference between the quasiparticles
and quasiholes, their generalized screened interactions $\kappa_{q\sigma r\tau}$,
the repulsion among quasiparticles and among quasiholes, and the attraction
between quasiparticles and quasiholes. 

To understand Eqs. (\ref{eq:QPpm}), consider the case
of just one excited quasiparticle, meaning all $\delta n_{q\sigma}=0$,
except one unoccupied state $\delta n_{a\sigma}=1$. Eq. (\ref{eq:OE_interpretation})
is then the excitation energy of adding an electron to the state $\phi_{a\sigma}$, 

\begin{equation}
\Delta E_v(p)=\varepsilon_{a\sigma}+\tfrac{1}{2}\kappa_{a\sigma,a\sigma},
\end{equation}
which is exactly the ground state quasiparticle energy $\varepsilon_{a\sigma}^{QP}$
based on a quadratic approximation, Eq. (\ref{eq:OE_interpretation}),
to the \ensuremath{\Delta}SCF energy and is consistent with Eq. (\ref{eq:QPpm}).
Similarly, consideration of one quasihole in Eq. (\ref{eq:OE_interpretation})
leads to the excitation energy of removing an electron from orbital
$\phi_{i\sigma}$,
\begin{equation}
\Delta E(h)=-\varepsilon_{i\sigma}+\tfrac{1}{2}\kappa_{i\sigma,i\sigma},
\end{equation}
which is exactly the ground state quasihole energy $\varepsilon_{i\sigma}^{QP}$
based on a quadratic approximation, Eq. (\ref{eq:OE_interpretation}),
to the \ensuremath{\Delta}SCF energy and is consistent with Eq. (\ref{eq:QPpm}).


The ground quasiparticle energies defined in Eq. (\ref{eq:QPpm}) are also exactly the expressions from the delocalization correction method GSC2 developed in the Yang laboratory  \cite{meiExactSecondOrderCorrections2021a,zhengImprovingBandGap2011}, and are much better approximation to the experimental ionization and electron affinity, than the orbital energies. 

Note that for the exact functional, the total energy as a function of fractional
charges connecting between $N$-electron system to $(N\pm1)$ systems,
has been proved to be linear, from ground state to ground states\cite{perdewDensityFunctionalTheoryFractional1982a, yangDegenerateGroundStates2000} and from ground state or excited states to excited states\cite{yangFractionalChargesLinear2024b}. Thus the chemical potentials, the
slopes of the curves are constant, and so are orbital energies, which
are the corresponding chemical potentials \cite{yangOrbitalEnergiesAre2024}.


In terms of quasiparticles and quasiholes, the excited state orbital energies, Eq. (\ref{eq:ExOrbEnergy}) become 
\begin{equation}
\text{\ensuremath{\varepsilon_{i\sigma}^{*}}}=\varepsilon_{i\sigma}-\sum_{j}^{N_{h}}\kappa_{i\sigma j\tau}+\sum_{a}^{N_{p}}\kappa_{i\sigma a\tau},\label{eq:QPminusGeneral}
\end{equation}
for orbitals corresponding to the ground state occupied orbitals,
and 
\begin{equation}
\text{\ensuremath{\varepsilon_{a\sigma}^{*}}}=\varepsilon_{a\sigma}+\sum_{b}^{N_{p}}\kappa_{a\sigma b\tau}-\sum_{i}^{N_{h}}\kappa_{a\sigma i\tau},
\label{eq:QPplusGeneral}
\end{equation}
for orbitals corresponding to the ground state unoccupied orbitals.
Thus the excited state orbital energies are modified from those of
the ground state by the quasiparticle interactions in the excited
state. 

The simplest OE expansion with 1 quasiparticle or 1 quasihole state, OE(p) or OE(h), has already been realized in our QE-DFT method -- using the DFA orbital energies to approximate quasiparticle
energies \cite{meiApproximatingQuasiparticleExcitation2018}, corresponding to the first term in Eqs. (\ref{eq:QPpm}). Correcting delocalization error in DFAs with GSC2 includes the second order term with better accuracy\cite{meiExactSecondOrderCorrections2021a,zhengImprovingBandGap2011}.

The next interesting case is OE(ph), with the 1 quasiparticle \textit{and} 1 quasihole state; its  total energy in Eq. (\ref{eq:OE_interpretation}) becomes
\begin{equation}
\text{\ensuremath{\Delta E(ph)=\varepsilon_{a\sigma}^{QP}-\varepsilon_{i\tau}^{QP}-\kappa_{a\sigma,i\tau}}},\label{eq:E1p1h}
\end{equation}
clearly showing the optical gap as the sum of the quasiparticle
energy gap and the particle-hole attraction. Extensive numerical study will be presented. Note that for charge transfer excitation, Eq.(\ref{eq:E1p1h}) has the correct long range Coulomb attraction.

We now compare our OE method with previous approaches. 

(1) For homogeneous systems, the LFL theory has a similar Taylor expansion, Eq. (\ref{LFL}). However, it has not been used to study individual excitation
energies\cite{chakravartyDynamicResponseInhomogeneous1979,LandauFermiLiquidTheory1991,nozieres1999theory}. Note that LFL theory already has the
quasiparticle energies for excited states, Eq. (\ref{eq:ExOrbEnergy}).
For inhomogeneous systems, the LFL theory involves expansions
with respect to the one-particle density matrix \cite{nozieres1999theory,LandauFermiLiquidTheory1991}.
In contrast, OE performs the expansion using orbital occupations.
Therefore, the expansion and interaction terms $\kappa$ in OE are fundamentally
different from the interaction parameters in LFL theory.

(2) Slater and coworkers have developed a Taylor expansion for ionization
or excitation energy within the $X_{\alpha}$ approximate functional
\cite{slaterStatisticalExchangeTotal1970c,slaterStatisticalExchangeCorrelationSelfConsistent1972a},
very similar to our OE expansion, Eq. (\ref{eq:OE_Energy}). All the
Taylor expansion coefficients, the partial derivatives, in Slater's
work have been calculated from fitting the Taylor expansion to the
numerical SCF calculations at various fractional occupations, leading
to good ionization energies \cite{slaterStatisticalExchangeTotal1970c,slaterStatisticalExchangeCorrelationSelfConsistent1972a}.
In addition, Slater's transition state method uses SCF calculations
at half occupations to obtain the excitation energies at integer occupations
without higher derivatives \cite{slaterStatisticalExchangeTotal1970c,slaterStatisticalExchangeCorrelationSelfConsistent1972a}.
In all cases, additional \ensuremath{\Delta}SCF excited state calculations
are used. Therefore, Slater's work cannot be used for the extrapolation
from ground state calculations to the excitation energies, which is
the main advance of our OE method. 

(3) Compared with the BSE approach for 1p1h excitation energy, Eq.
(\ref{eq:E1p1h}) has very similar physics involving the gap of quasiparticle
energies and the screened interaction. The key difference is that our quasiparticle
energies are all defined within DFT and our generalized screened interaction $\kappa $ \cite{meiExactSecondOrderCorrections2021a} 
can generally be a four point quantity and include  exchange
and correlation contributions beyond the Coulomb interactions screened with linear response based on the random phase approximation used in BSE. Fundamentally, OE approximates \ensuremath{\Delta}SCF,
which is based on a single determinant, while BSE uses multi-determinants. 

Indeed, OE inherits the limitations of single-determinant \ensuremath{\Delta}SCF
approaches. In particular, states with substantial multi-configurational
character are not well described. Moreover, the spin symmetry breaking
may occur when an excited singlet state becomes open-shell, leading
to spin contamination from the $m_{s}=0$ triplet solution. The same
issue exists for ground state DFT, but the scenarios, such with open-shell
singlets, are more common for excited states. To correct this issue, we employ a widely used spin-purification scheme
\cite{bagusSingletTripletSplittings1975,zieglerCalculationMultipletEnergies1977},
assuming that the mixed state is primarily contaminated by a single
triplet component and that $\langle S^{2}\rangle$ is close to 1.
The purified singlet energy is obtained as 
\begin{equation}
E_{S}=2E_{\text{B.S.}}-E_{T},
\end{equation}
where $E_{\text{B.S.}}$ is the spin-contaminated (broken symmetry)
state energy, and $E_{T}$ is the triplet energy obtained from the
$m_{s}=\pm1$ solutions. This correction has been shown to effectively
remove spin contamination in \dscf calculations \cite{kowalczykAssessmentDSCFDensity2011},
and all singlet excitation energies reported in this letter include
this purification.

Other theoretical extensions of OE are possible. We can go beyond
the quadratic form and use higher-order derivatives, which are available
\cite{pengFukuiFunctionResponse2013}. So far, we have focused on OE
expansion for excitation energy from the neutral ground states, but
we can use any other single-determinant reference systems to make
the OE expansion. 

Computationally, for the exchange-correlation component in $K$ obtained
from local and semi-local functionals, we use a Sherman-Morrison-based 
Resolution of the Identity approximation to decrease the computational
cost to $\mathcal{O}(N^{3})$ \cite{lrLOSCValence}. 


We first demonstrate that the OE expansion accurately predicts 1p1h
excitation energies across a diverse set of molecular excitations
using only derivatives obtained from a single ground-state calculation.
To evaluate its accuracy, we perform numerical benchmarks with the
augmented correlation-consistent triple-$\zeta$ basis set
(aug-cc-pVTZ) \cite{dunningGaussianBasisSets1989,kendallElectronAffinitiesFirstrow1992}.
OE is implemented in both the $\text{QM}^{4}\text{D}$ package
\cite{QM4D} and PySCF \cite{sunPySCFPythonbasedSimulations2018}.
Unless otherwise noted, the results reported in the main text are based
on Kohn--Sham calculations using the BLYP functional
\cite{beckeDensityfunctionalExchangeenergyApproximation1988,leeDevelopmentColleSalvettiCorrelationenergy1988}
(OE@BLYP). Since OE can also be applied to generalized
Kohn--Sham references, we additionally assess OE using the B3LYP
functional \cite{beckeDensityfunctionalExchangeenergyApproximation1988,leeDevelopmentColleSalvettiCorrelationenergy1988,stephensInitioCalculationVibrational1994}
(OE@B3LYP); these results are provided in the SM\cite{SM}.
For comparison, $\Delta$SCF calculations are carried out in PySCF
using the maximum overlap method (MOM)
\cite{gilbertSelfConsistentFieldCalculations2008a} to obtain
excited-state solutions.

\begin{figure}
    \centering
    \includegraphics[width=\linewidth]{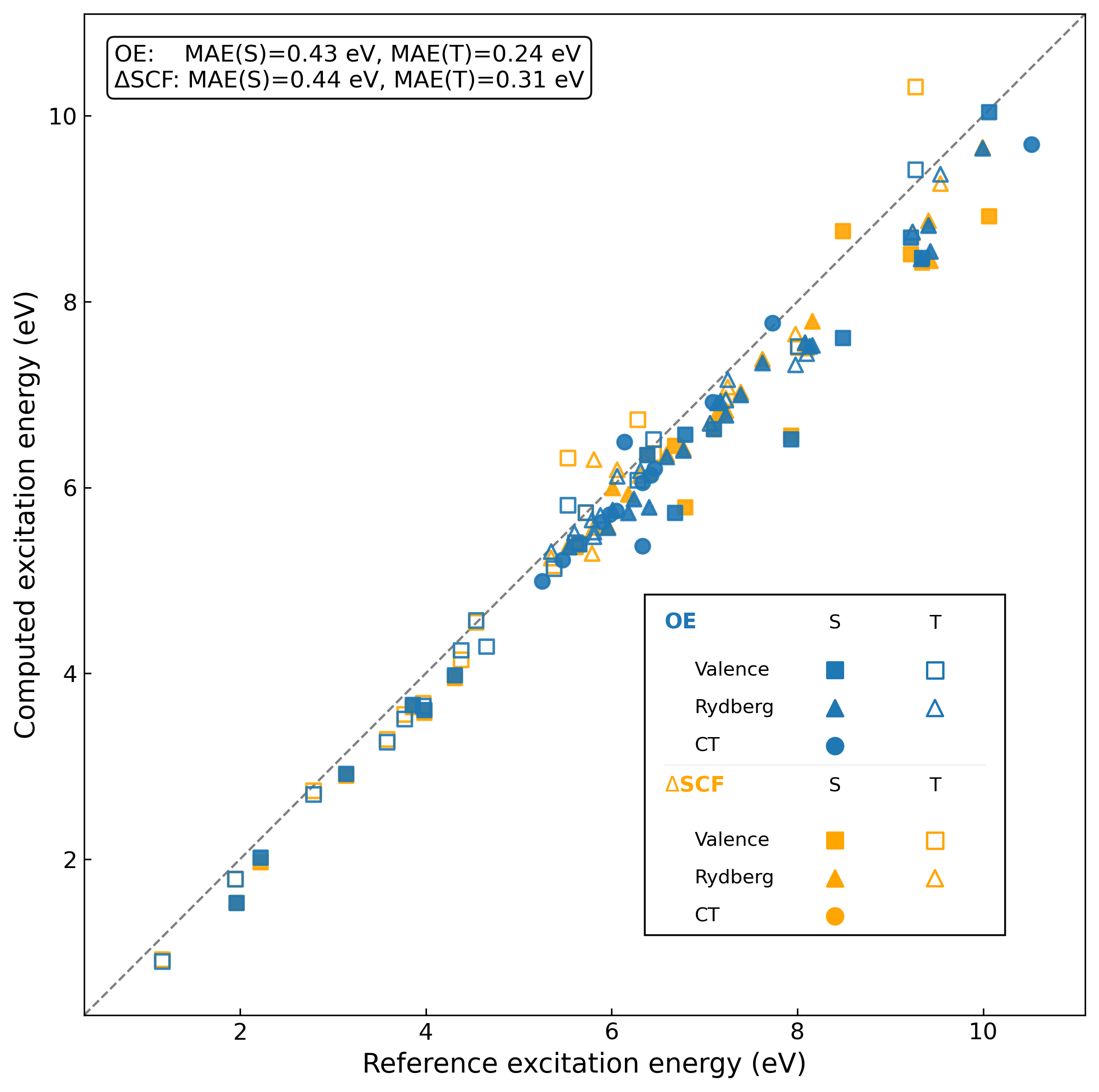}
    \caption{Comparison of excitation energies obtained from OE@BLYP (blue markers) and $\Delta$ BLYP (yellow markers) with theoretical best estimates (TBE). The dashed line indicates perfect agreement with the reference values. OE closely reproduces \dscf excitation energies across different excitation types, yielding comparable mean absolute errors (MAEs) for both singlet (S) and triplet (T) states.}
    \label{fig:scatter}
\end{figure}

Fig.~\ref{fig:scatter} compares excitation energies obtained from
OE@BLYP and $\Delta$BLYP with theoretical best estimates (TBE)
for three major excitation categories: valence excitations
\cite{loosMountaineeringStrategyExcited2018b}, Rydberg states
\cite{xuTestingNoncollinearSpinFlip2014, loosMountaineeringStrategyExcited2018b, NIST_ASD}, and charge-transfer (CT)
excitations \cite{kozmaNewBenchmarkSet2020a}. Reference values are
taken from high-level wave function methods (see SM\cite{SM} for more details).

Overall, OE shows excellent agreement with both $\Delta$SCF and the
reference values across all excitation types, with most data points
closely distributed around the diagonal. Quantitatively, for valence
excitations, OE@BLYP achieves mean absolute errors (MAEs) of
0.55~eV for singlets and 0.29~eV for triplets, comparable to, and in
some cases slightly smaller than, the corresponding $\Delta$BLYP
results. Similarly, for Rydberg excitations, OE@BLYP yields MAEs
of 0.44~eV for singlet states and 0.23~eV for triplets, closely
reproducing the $\Delta$BLYP values. These results confirm that OE
captures the energetic trends across different excitation classes
while faithfully reproducing the $\Delta$SCF excitation energies.

Charge-transfer (CT) excitations are known to challenge
standard density functionals lacking sufficient long-range
exact exchange, and they often lead to SCF convergence
difficulties in \dscf calculations employing the
maximum overlap method (MOM). TD-DFT with
range-separated hybrid functionals and system specific tuned
parameters is commonly used to address these issues.\cite{baerTunedRangeSeparatedHybrids2010a}
In contrast, OE provides an alternative route that does
not require tuning or long-range exact exchange.
Numerical results show that OE reproduces CT excitation
energies in close agreement with high-level
EOM-CCSDT-3 reference values\cite{kozmaNewBenchmarkSet2020a}, with a mean absolute
error of 0.30 eV.

The pure functional starting point in OE@BLYP shows good alignment with both $\Delta$BLYP and reference values reported in the literature. The local and semi-local exchange correlation kernel also enables an efficient RI-like approximation for the analytical second derivative. Nevertheless, the presence of delocalization error \cite{cohenInsightsCurrentLimitations2008, mori-sanchezLocalizationDelocalizationErrors2008} remains a notable source of error in calculations for both OE and $\Delta$BLYP. Implementing exact exchange can help mitigate the delocalization error. Further analysis using the B3LYP functional \cite{beckeDensityfunctionalExchangeenergyApproximation1988, leeDevelopmentColleSalvettiCorrelationenergy1988, stephensInitioCalculationVibrational1994} indicates a general enhancement in OE and \dscf across both valence and Rydberg excitations. For valence excitations, OE@B3LYP shows a mean absolute error (MAE) of 0.44 eV for singlets and 0.25 eV for triplets, whereas a significant enhancement was seen in Rydberg singlet excitations, with the MAE decreasing from 0.44 eV in OE@BLYP to 0.23 eV in OE@B3LYP. This enhancement indicates that delocalization error also greatly impacts Rydberg states, even in small molecules.

We also presented detailed numerical results in SM\cite{SM} for OE excited state orbital energies, Eqs. (\ref{eq:QPminusGeneral},\ref{eq:QPplusGeneral}), showing their agreement with the $\Delta$SCF calculations.

Despite OE demonstrating high accuracy across various types,
certain inherent challenges persist. 
A major source of error in OE originates from its reliance on Kohn--Sham
orbitals and densities obtained from approximate functionals, which in general have delocalization error\cite{cohenChallengesDensityFunctional2012}. The
OE energy expansion in Eq.(\ref{eq:OE_Energy}) assumes that \dscf
total energies at integer occupations are accurate. While this assumption
generally holds for small molecules near equilibrium geometries, it
breaks down in large, extended or stretched systems. For example, DE is known
to affect both ground and excited states in bulk materials \cite{cohenFractionalChargePerspective2008,cohenChallengesDensityFunctional2012}.
In addition, recent studies also reveal that conventional DFAs poorly describe
core orbitals, which impacts the accuracy of core-level excitation
and binding energies \cite{yuAccuratePredictionCoreLevel2025}. Incorporating correction to delocalization error, including the lrLOSC\cite{yuAccuratePredictionCoreLevel2025,lrLOSCValence}, into OE may
offer a path forward for handling core excitations and extended systems more reliably with minimal functional dependence.

Numerical challenges in OE calculations can also emerge.
In particular,  the
second-order derivatives of some local density functional approximations exhibit
singularities in low density regions for certain molecular calculations. In practice, some
unphysical negative curvatures are detected \cite{liPiecewiseConvexConcave2017},
which originated from the singularity in the bare exchange-correlation kernel. For instance, the Local Density Approximation (LDA) \cite{kohnSelfConsistentEquationsIncluding1965} functional is formulated as
$E^\text{LDA}_\text{x} [\rho(\mathbf{r})] = -C_\text{x}\int  \rho^{4/3}(\mathbf{r}) d\boldsymbol{r}$,
and its second functional derivative includes the negative power of electron density. The xc kernel contains large components in small density regions, which can significantly affect the integrals that only involve virtual states. In order to avoid such numerical failures, in this work, we use the analytical curvatures from a small fraction away from the integer point, with fixed orbitals at the integer occupation and without any additional SCF calculation for orbital relaxation. The details of the method and its validation are presented in SM\cite{SM}. This strategy regularizes singularities from the xc functional derivatives, and for consistency, all numerical results reported here are obtained using this regularization. 

In conclusion, this study presents Occupancy Extrapolation (OE) as an efficient, accurate, and flexible approach that captures many ionization, electron addition and excitation energies from a single ground-state DFT calculation. OE provides the needed analytical  energy derivatives for the Taylor expansion and accurately replicates \dscf excitation energy across valence, Rydberg, and charge-transfer states without any state-specific SCF calculations. Further developments incorporating delocalization error correction for DFA and higher-order Taylor expansion should further enhance the accuracy and broaden its applicability.

We acknowledge support from the National Institute of Health (1R35GM158181-01).

\bibliography{OE_abbr}

\end{document}